# Crystal melting influenced by particle cooperativity of the liquid

P. Lunkenheimer[1,]*, K. Samwer[2], and A. Loidl[1]
[1]Experimental Physics V, Center for Electronic Correlations and Magnetism, University of Augsburg, 86159 Augsburg, Germany
[2]1. Physikalisches Institut, University of Göttingen, 37077 Göttingen, Germany

Recently, a universal relation between the thermal expansion coefficient of glasses $\alpha_g$, their glass-transition temperature $T_g$, and the so-called fragility index $m$ of the corresponding supercooled liquid state was found to be valid for more than 200 glass formers, namely $\alpha_g/m \propto 1/T_g$ [P. Lunkenheimer *et al.*, Nat. Phys. **19**, 694 (2023)]. Here we show that this could also have far-reaching consequences for our understanding of crystal melting. Namely, when considering the empirically founded 2/3-rule, stating that the ratio of $T_g$ and the melting temperature $T_m$ is about 2/3 for almost all materials, for crystals a similar relation, $\alpha_c/m \propto 1/T_m$, should apply. Indeed, we find that the available experimental data are well consistent with such a relation. This implies that the melting of a crystal into an ordinary (non-supercooled) liquid is influenced by the fragility, a property quantifying the non-Arrhenius dynamics in the supercooled-liquid state of the material. We argue that this can be explained by a significant enhancement of the "ideal" (non-cooperative) melting temperature arising from the cooperativity of the particle motion in the liquid state above $T_m$. Therefore, a reassessment of the currently widely accepted microscopic understanding of crystal melting, still founded on the general ideas that lead to the time-honored Lindemann melting criterion, may be necessary.

## I. INTRODUCTION

The Lindemann criterion for the melting of a crystal into a liquid [1,2] is a well-established concept in condensed matter physics [3]. It essentially states that crystalline materials melt when the particle displacements caused by thermal vibrations exceed a certain percentage (roughly 10-20 % [4,5,6]) of their average lattice-site spacing. While the original idea goes back to the kinetic theory of solids by Sutherland [7], one should be aware that a clear-cut theoretical derivation of the Lindemann criterion is still missing, and it should be regarded as semi-empirical. As explained, e.g., in Ref. [8], based on the reasoning behind the Lindemann criterion, a correlation of the melting temperature $T_m$ with the thermal expansion coefficient $\alpha_c$ of the crystal can be expected, namely [5,9]:

$$\alpha_c \propto 1/T_m \tag{1}$$

As discussed in Ref. [8], if $U_0$ is the depth of the pair-potential well, whose asymmetry gives rise to thermal expansion, the inverse proportionality of Eq. (1) is based on the reasonable assumptions that $T_m \propto U_0$ (with $U_0$ the depth of the well) [10,11,12] and $1/\alpha_c \propto U_0$ [13]. Within distinct classes of crystalline materials, the approximate validity of Eq. (1) was indeed confirmed experimentally [5,14,15].

Aside of crystallization, a qualitatively different path towards solidification, in principle available to almost any liquid [16], is its supercooling and final kinetic arrest into a glass, a solid state lacking the periodicity of a crystalline lattice [17,18,19,20]. Supercooling is achieved by cooling a liquid sufficiently fast to avoid crystallization at $T_m$, the most common way to produce a glass. (Other procedures are also possible, e.g., strain-driven glass transitions [21].) Below the glass-transition temperature $T_g$, then the particle dynamics becomes so slow (and the viscosity so high) that the resulting glass can be considered as solid for all practical purposes. $T_g$ is usually defined as the temperature where the viscosity $\eta$ exceeds $10^{12}$ Pa·s or where the relaxation time $\tau$, characterizing particle mobility, exceeds 100 s (according to the Maxwell relation, both quantities are approximately proportional to each other). Numerous competing theories were proposed to explain this so-called glass transition, whose microscopic nature thus still can be considered as controversial. At first glance, it reminds of a second-order phase transition, because quantities like the specific heat and thermal expansion exhibit jumplike (whatsoever, rather smeared out) behavior when crossing $T_g$. However, the fact that this liquid-glass crossover depends on the cooling rate, rules out that $T_g$ simply marks a canonical phase transition. Instead, it is clear that the material falls out of thermodynamic equilibrium when cooling below $T_g$. This is due to the continuous slowing down of the dynamics of the particles, preventing their proper rearrangement into equilibrium positions before the temperature has further fallen [18,19]. Consequently, scenarios were proposed where the glass transition is seen as purely dynamic phenomenon, without invoking any phase transition [22,23,24,25]. However, there are also various models that instead assume an underlying "ideal" phase transition at a temperature below [26,27,28,29] or above $T_g$ [30,31]. This helps to explain the typical noncanonical properties of the supercooled-liquid state, the most prominent one being the non-Arrhenius temperature dependence of $\eta$ and $\tau$. Unfortunately, due to the inevitable falling out of equilibrium upon cooling below $T_g$, this suggested ideal glass transition cannot be experimentally accessed for any reasonable cooling rate. However, based on theoretical advances [32,33], especially recent experiments measuring higher-order susceptibilities seem to support such a "hidden" phase transition [34,35,36].

In a recent work by the present authors and collaborators [8], the question was raised whether a Lindemann-like criterion may also govern the solid-liquid transition of glasses at $T_g$. Such a notion was earlier considered, e.g., in Refs. [5,37,38,39,40,41,42]. To help clarify this question, for more than 200 glass-forming materials, the corresponding relation to Eq. (1),

$$\alpha_g \propto 1/T_g, \tag{2}$$

was checked (with $\alpha_g$ the expansion coefficient in the glass state). They all belonged to very different material classes: molecular liquids, polymers, ionic systems like ionic liquids and melts, metals, and network glass formers, the latter including silicate glasses as used in everyday life for windows, bottles, etc. A clear failure of this proportionality was found. However, interestingly it was noted that a scaling of $\alpha_g$ with the so-called fragility index $m$ can restore this proportionality, namely the relation

$$\alpha_g/m \propto 1/T_g \tag{3}$$

was found to be valid [8]. The fragility index was introduced in Refs. [43,44,45] to quantify the degree of deviation of $\eta(T)$ or $\tau(T)$ of glass-forming liquids from simple thermally-activated temperature dependence. The latter should lead to an Arrhenius law, $\eta$ or $\tau \propto \exp[E/(k_B T)]$ (where $E$ is the energy barrier). However, instead a stronger temperature dependence (sometimes termed "super-Arrhenius") is commonly found in glass formers. It can be often reasonably parameterized [17,18,19,20,46,47,48] by the empirical Vogel-Fulcher-Tammann (VFT) formula,

$$\tau = \tau_0 \exp\left(\frac{DT_{VF}}{T - T_{VF}}\right) \tag{4}$$

(or the corresponding equation for $\eta$) [49,50,51,52]. Here $1/(2\pi\tau_0)$ is an attempt frequency, typically of the order of a phonon frequency. The divergence temperature $T_{VF}$ may be regarded as an estimate of the mentioned underlying phase-transition temperature, but one should be aware that also alternative formulae can describe the experimental data, not involving any divergence temperature (see, e.g., Refs. [46,48]). This includes $\tau(T)$ as predicted by the generalized entropy theory of glass-formation [53]. Here we merely employ Eq. (4) as an empirical, often-used formula to approximately parameterize $\tau(T)$ or $\eta(T)$ in the whole temperature range above $T_g$. The conclusions of the present work do not rely on the assumption of a relaxation-time or viscosity divergence. The strength parameter $D$ in Eq. (4) [52] determines the deviations from Arrhenius temperature dependence, just as the more commonly used fragility index $m$, mentioned above.

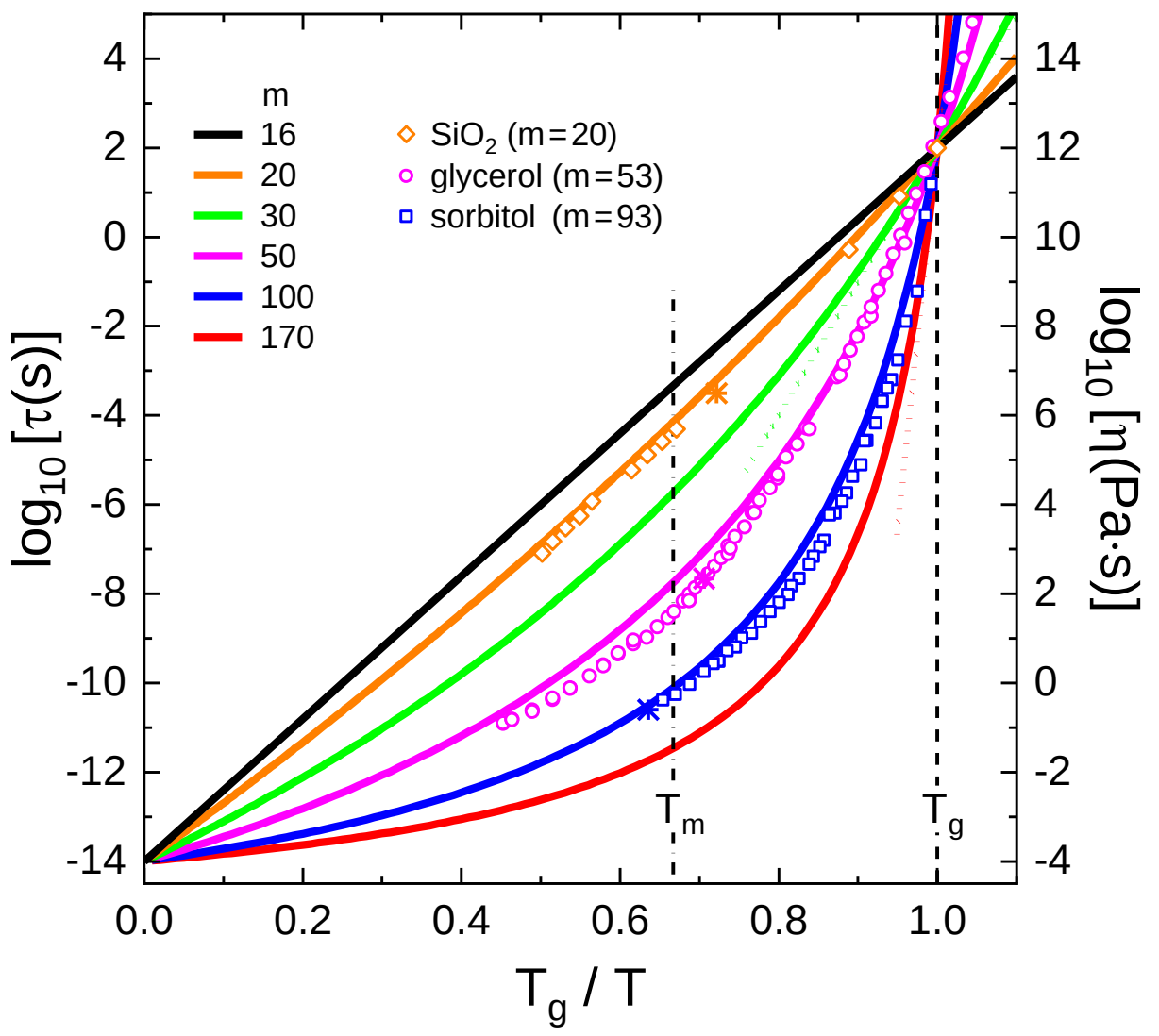

FIG. 1. Angell plot of the temperature-dependent relaxation time (left ordinate) and viscosity (right ordinate). The solid lines calculated using the VFT formula, Eq. (4), schematically illustrate the different behavior in dependence of the fragility for values of $m$ between 16 and 170. The slope at $T_g$, exemplarily indicated by the dotted lines for $m$ = 30 and 170, defines the fragility index $m$ [43,44]. The open symbols show three experimental examples ($\tau$ of glycerol [46], $\tau$ of sorbitol [47], and $\eta$ of $SiO_2$ [52]) with different $m$ values as indicated in the right figure legend [45]. The stars show the respective experimental melting temperatures [15,65,70]. The vertical dashed and dash-dotted lines indicate $T_g$ and $T_m \approx 3/2\ T_g$, respectively.

The solid lines in Fig. 1 show typical VFT curves calculated from Eq. (4) in an Angell plot [54], $\log_{10}(\tau)$ or $\log_{10}(\eta)$ vs $T_g/T$ (left and right ordinates, respectively). Within this representation, the fragility index $m$ is defined by the slope at $T_g$ [43,44,45]. The steeper this slope (dotted lines in Fig. 1, exemplarily shown for two $m$ values), the more $\eta(T)$ or $\tau(T)$ deviate from the Arrhenius law, which appears as straight line in this type of plot. Glass formers where these deviations are well-pronounced are termed "fragile" and those where they are weak are denoted "strong" [52]. Overall, the fragility is an important quantity in glass physics and many properties of glass formers were found to correlate with $m$ (see, e.g., [17,45,55,56]). Assuming $\tau_0 \approx 10^{-14}$ s and $\tau(T_g) \approx 100$ s [57], pure Arrhenius behavior corresponds to $m \approx 16$ [58]. Under the same assumptions, the fragility index can also be calculated from the VFT parameter $D$, via $m \approx 16 + 590/D$ [45]. As typical examples, the open symbols in Fig. 1 represent experimental data for three glass formers with different fragilities. In the network glass-former $SiO_2$ ($m$ = 20 [45]), $\eta(T)$ [52] nearly follows Arrhenius behavior – it is a strong system. For the two molecular supercooled liquids glycerol and sorbitol, $\tau(T)$ is shown [46,47]. Sorbitol ($m$ = 93 [45]) can be classified as fragile, while $\tau(T)$ of glycerol ($m$ = 53 [45]) reveals intermediate characteristics [52].

An often-assumed explanation of the universal super-Arrhenius behavior of glass formers is increasing cooperativity of the particle motion when the glass transition is approached upon cooling [18,19,27,28]. This leads to an increasing length scale of cooperatively rearranging regions (CRRs), originally proposed in the Adam-Gibbs theory of the glass transition [27]. Such a scenario was recently corroborated by measurements of nonlinear susceptibilities, detecting the growth of CRR sizes upon cooling in various glass formers, which is most pronounced in fragile ones [34,35,36,59]. Within this framework, the empirically found relation, Eq. (3), was proposed to arise from an enhancement of the glass-transition temperature for fragile systems, compared to a value that would be detected in the absence of cooperativity [8]. This was based on the reasonable assumption that for these glasses more energy is needed to break up their extended CRRs. In Ref. [8] it was suggested that then an additional factor $m$ should be introduced into the relation $T_g \propto U_0$, leading to $T_g \propto m\, U_0$, thus enhancing $T_g$ by a cooperativity-dependent factor. [More precisely, the enhancement factor can be assumed to be $m/16$, implying no cooperativity-induced increase for strong glasses, but the 1/16 factor can be regarded as part of the proportionality factor in Eq. (3).] Together with $1/\alpha_c \propto U_0$ [13], this rationalizes the empirically found validity of Eq. (3) [8]. Interestingly, molecular dynamics simulations of polymer melts using a "bead-spring" model have revealed a decrease of the fragility and an increase of $T_g$ with increasing strength of the attractive bead interactions [60,61,62]. This is consistent with the relation $T_g \propto m\, U_0$ considered in Ref. [8]. Finally, we want to note that the generalized entropy theory of glass-formation [63] predicts an increase of fragility with the product of $\alpha_g$ and $T_g$, in accord with Eq. (3). This theory also considers the cooperative nature of the glass transition as discussed above and may provide a theoretical basis for the validity of Eq. (3).

## II. THE 2/3 RULE

One should note that the thermal expansion in the glass and crystal state is dominated by the same process, namely local vibrations within the anharmonic interparticle potential. The latter is essentially the same for both states, reflecting their similar short-range order, and, thus, $\alpha_g$ and $\alpha_c$ should be nearly identical [18,20,64]. However, a severe problem arises from the above considerations: As already remarked in Ref. [8], then the assumption of the validity of both Eqs. (1) and (3), leads to a clear contradiction to the often-assumed, quite universal relation [15,19,37,64,65,66,67,68,69],

$$T_g = 2/3\, T_m \,, \qquad (5)$$

known in glass physics as "2/3 rule". The validity of Eq. (3) is well established by the very broad data set in Ref. [8]. Therefore, either Eq. (1) or (5) should be invalid. In the following, we first check the latter.

The vertical dash-dotted line in Fig. 1 indicates $T_m$ as expected according to Eq. (5). The actual melting temperatures of the three included glass formers (stars [15,65,70]) lie within the vicinity of this line, which points to the approximate validity of Eq. (5). For a more thorough check, in Fig. 2 we present $T_g$ vs $T_m$ for more than 100 glass formers, mainly concentrating on those already analyzed in Ref. [8] and on such materials where thermal-expansion data are available for the crystalline state to be used in the analysis below. A list of the used data is provided in Table SI in the Supplemental Material [71] (including references [8,15,65,66,70,72–123]). The main frame of Fig. 2 shows these data in double-logarithmic representation. The line represents a linear fit with slope 1, leading to a good description of the experimental data, which points to direct proportionality of the two temperatures. The obtained proportionality factor of 0.65 is reasonably close to the often-assumed value of 2/3 in Eq. (5) (for $T_g$ values below ~60 K, not considered here, quantum effects can lead to deviations [123]). The inset of Fig. 2 shows the same data in linear representation, directly visualizing the linear relation between $T_g$ and $T_m$ with zero intercept and slope ~2/3. Overall, in accord with earlier findings [65,66,68], Eq. (5) can be considered as approximately valid, although, to our knowledge, it lacks a clear-cut theoretical explanation up to now.

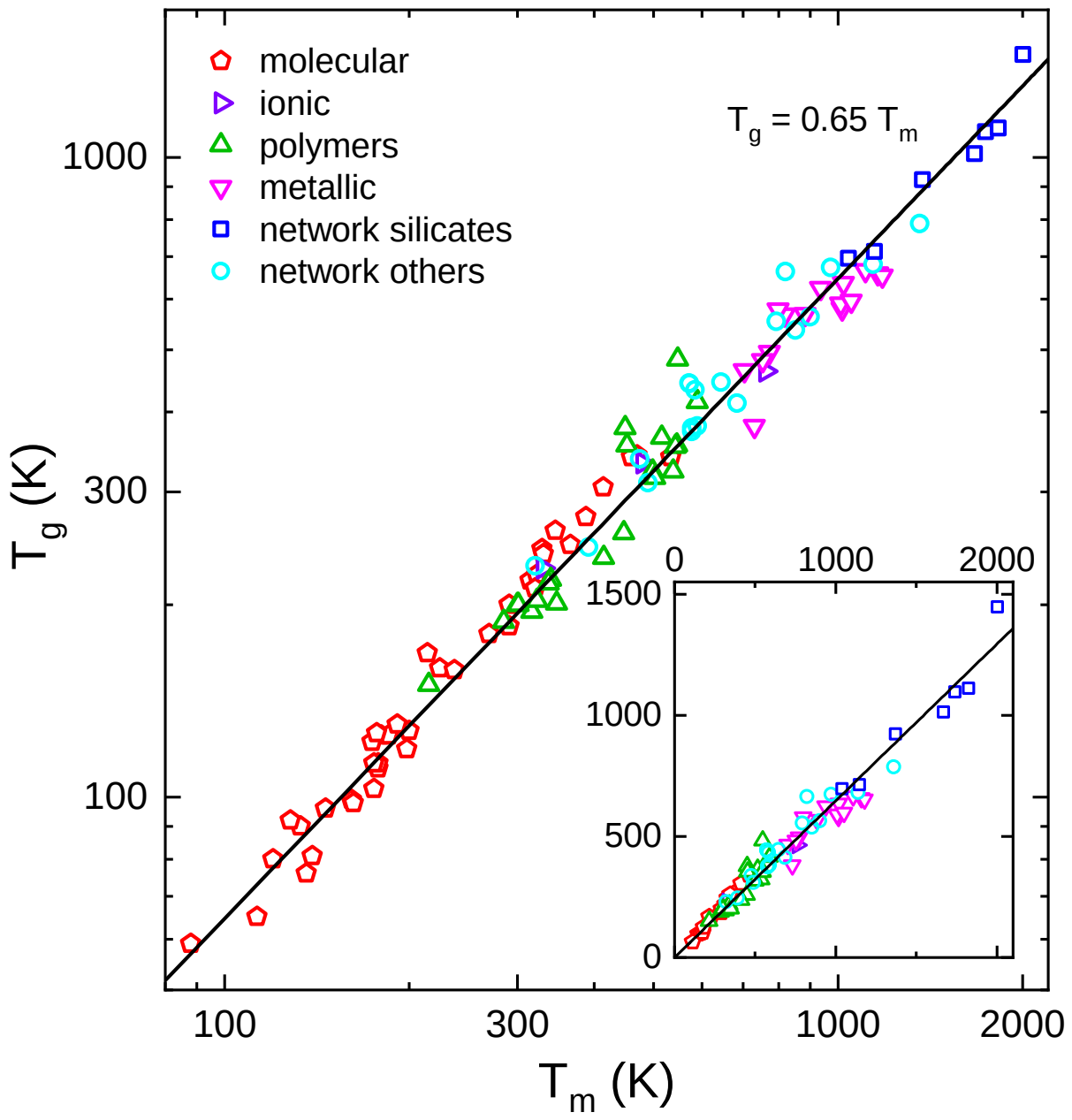

FIG. 2. Double-logarithmic plot of $T_g$ vs $T_m$ for more than 80 glass formers belonging to different material classes as indicated in the legend (see Table SI [71] for a list of all data points and sources). The line is a fit with $T_g \propto T_m$ (corresponding to a straight line with slope one in this representation), leading to a proportionality factor of 0.65. The inset shows the same data in linear representation.

While the average of the experimentally found $T_g/T_m$ values is close to 2/3 [65,66,68], the actual experimental values can vary between about 0.5 and 0.8, as shown, e.g., for polymers in Ref. [124]. If both Eqs. (1) and (3) would be valid, one would arrive at $T_g/T_m \propto m$. However, depending on the material, $m$ can vary between about 20 and 170 [68], a factor of 8.5. Thus, for different materials $T_g/T_m$ should vary by this factor, too. In contrast, as mentioned above, the experimental values for $T_g/T_m$ roughly vary between 0.5 and 0.8 [65,123,124,125], i.e., by a significantly smaller factor of about 1.6. Moreover, there is no indication for a systematic variation of $T_g/T_m$ with $m$. Therefore, the conclusion in the beginning of this chapter, that the simultaneous validity of Eqs. (1) and (3) is excluded, remains correct: In light of Eq. (5), they cannot both be valid, even when considering the observed scatter in the 2/3 value.

## III. INFLUENCE OF FRAGILITY ON CRYSTAL MELTING

As, thus, Eqs. (3) [8] and (5) (Fig. 2) are experimentally well founded, the above-discussed inconsistency of Eqs. (1), (3), and (5) can only be resolved when rejecting Eq. (1). As mentioned above, its validity was checked within different materials classes [5,14,15], but not across a similarly broad collection of materials as done for Eq. (2) (found to be invalid) and (3) (valid) in Ref. [8]. The simplest solution would be to apply a similar fragility scaling to Eq. (1) as it was done for the glass case, leading to the modification of Eq. (2) into Eq. (3). To illustrate the latter, Fig. 3(a) shows the effect of fragility scaling on the $T_g$-dependent expansivity of glasses as treated in detail in Ref. [8] [compared to Fig. 1(e) of that work, some additional data points are included in Fig. 3(a), especially for metallic glasses; see Table SII [71]]. The bare $\alpha_g$ (open symbols) decreases significantly stronger with $T_g$ than expected from Eq. (2) and can be roughly fitted by $\alpha_g \propto 1/T_g^{-2.2}$ (dashed line) [126]. However, plotting instead $\alpha_g/m$ (closed symbols) leads to clear $1/T_g$ dependence, i.e., Eq. (3) is well fulfilled (the only exception is $SiO_2$ which has the smallest $\alpha_g$ and highest $T_g$ and reveals an anomalous density temperature-dependence [127]).

As mentioned in section I, the introduction of $m$ into Eq. (3) was motivated by an assumed enhancement of $T_g$ due to the particle cooperativity, which is most pronounced in fragile glass formers and should raise the energy needed to liquify a glass [8]. Could such a scenario indeed also apply to crystal melting? It would lead to

$$\alpha_c/m \propto 1/T_m \, , \qquad (6)$$

which, in contrast to Eq. (1), is compatible with Eqs. (3) and (5) when considering that $\alpha_g \approx \alpha_c$. To check the possible validity of Eq. (6), thermal-expansion data of such crystalline materials are needed, for which also dynamic data in their supercooled-liquid state are available, allowing for the determination of the fragility [e.g., from Angell plots or from VFT fits of $\tau(T)$ or $\eta(T)$]. Unfortunately, this requirement restricts the number of available data points that can be found in literature. The open symbols in Fig. 3(b) show $\alpha_c(T_m)$ data (Table SIII [71]) for 25 such systems belonging to different material classes as indicated in the legend in frame (a). They reveal a clear trend to stronger temperature dependence than suggested by Eq. (1), which was derived from the Lindemann criterion. As shown by the dashed line, a free power-law fit leads to $\alpha_c \propto 1/T_m^{-1.5}$ instead of $1/T_m$. In contrast, when plotting $\alpha_c/m$ [closed symbols in Fig. 3(b)], in accord with Eq. (6), this too strong temperature dependence becomes reduced, and the data points can be reasonably described by a $1/T_m$ behavior (solid line). An alternative fit of these data with $\alpha_c/m \propto 1/T_m^{-s}$ with free exponent $s$ (not shown) leads to $s = 0.95$, i.e., with negligible deviation from $s = 1$ presumed in Eq. (6). We conclude that the thermal-expansion data of the crystal state shown in Fig. 3(b) are well compatible with Eq. (6).

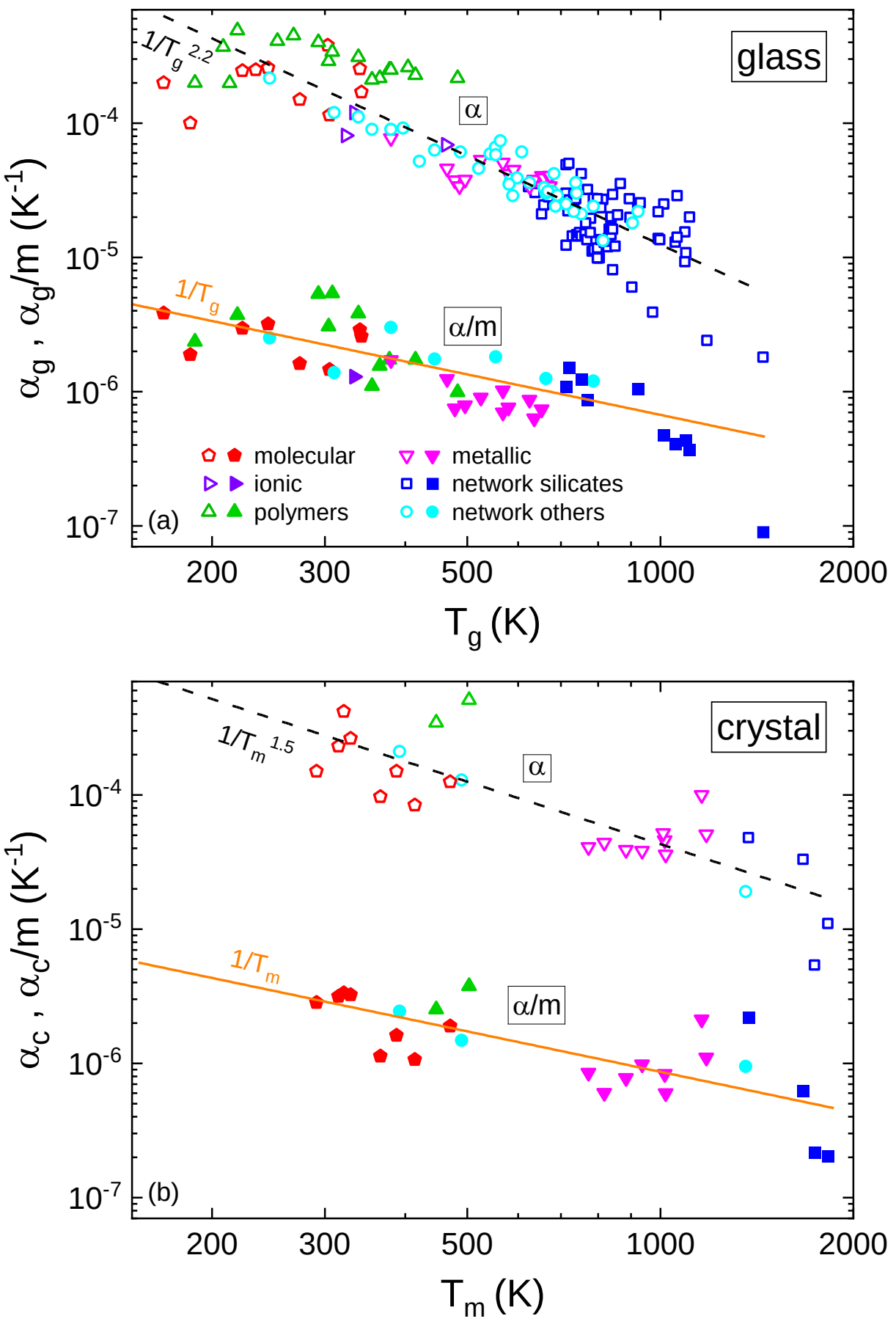

FIG. 3. Volume thermal-expansion coefficients $\alpha_g$ in the glass phase (a) [8] and $\alpha_c$ in the crystalline phase (b) (see Table SIII [71] for a list of all data points and sources), plotted double-logarithmically vs $T_g$ or $T_m$, respectively. The data cover a large variety of materials from different material classes as indicated in the legend. The open symbols represent the bare expansion coefficients while the closed symbols show $\alpha$ divided by $m$. The dashed lines are power-law fits, of the bare $\alpha$ data (open symbols), leading to exponents of about -2.2 for $\alpha_g$ and -1.5 for $\alpha_c$. The solid lines are fits of the $\alpha/m$ data with slope -1, corresponding to Eqs. (3) and (6). To facilitate a comparison of the $\alpha_g$ and $\alpha_c$ data, the ordinates and abscissae of both frames cover identical ranges.

## IV. DISCUSSION

At this point, a note of caution seems advisable. Like the $\alpha_g$ data analyzed in Ref. [8] [cf. Fig. 3(a)], the scatter of the data in Fig. 3(b) is considerable. However, in that work significantly more data points than in the present study were available, which largely compensated the uncertainties of the individual points and enhanced the significance of the found correlations. We refer the reader to the supplementary information of Ref. [8], where various sources of error were discussed in detail (e.g., the use of different experimental techniques, evaluation methods, etc.), which also applies for the present data. For these reasons, the results of Fig. 3(b), although based on data for 25 different glass formers, only can be regarded as a clear hint at the validity of Eq. (6) but not a definite proof. For such a proof, more experimental work on materials in both their crystalline and supercooled state is necessary. This is highly desirable because the possible validity of Eq. (6) has interesting consequences for such a fundamental process as crystal melting. Namely, this relation implies that a property known to govern the supercooled-liquid state, the fragility, plays a major role in the melting of the crystalline state.

To explain Eq. (6), in analogy to the reasoning for the glass transition mentioned in section I [8], $T_m \propto mU_0$ should be valid instead of $T_m \propto U_0$. Consequently, the melting temperature of fragile systems becomes enhanced (most likely by a factor $m/16$, see remark in section I) due to the cooperativity of the liquid, and, without cooperativity, $T_m$ would be significantly lower. This is surprising because the fragility $m$ is a quantity that by definition [45] is determined deep in the supercooled state, close to $T_g$ (cf. dotted lines in Fig. 1). Is it reasonable that the crystal somehow "knows" the degree of cooperativity of the material's supercooled-liquid state close to $T_g$? And is it possible that its melting is influenced by this property, although above $T_m \approx 3/2\, T_g$ the material transfers into a normal liquid, which is not supercooled at all? For the glass, a corresponding scenario is rather plausible, because the glass transition occurs at $T_g$, where $m$ is determined, and the structure of the glass is essentially the same as that of the supercooled liquid just above $T_g$. In the crystal, however, the structures of solid and liquid are different, although the short-range order in both phases usually is similar.

These concerns can be at least partly relieved when considering Fig. 1. Indeed, $m$ is determined at $T_g$ (dotted lines), but by no means the fragility of a glass former is a quantity that solely affects a liquid in its supercooled state. The curves drawn for different fragilities in Fig. 1 significantly deviate from each other, not only in the supercooled but also in the normal liquid state, even at lowest viscosities or smallest relaxation times approached for $T \rightarrow \infty$ [128]. This is also reflected by an alternative quantification of fragility, proposed by Richert and Angell [129], based on the value of $T_g/T$ at $\tau(10^{-6}\,\mathrm{s})$, which encompasses the liquid region for strong systems (cf. Fig. 1). Finally, according to Refs. [25] and [48], the fragility index seems to be connected to the softness parameter of the repulsive part of the pair potential, which is relevant for all phases, no matter whether crystal, liquid, or supercooled liquid [130].

For these reasons, fragility should be regarded as a property of every liquid, whether supercooled or not, and it can be expected to strongly influence its properties, also at high temperatures. This property is widely unknown outside glass physics because the degree of non-Arrhenius behavior of $\tau(T)$ or $\eta(T)$ can be best detected in liquids that can be easily supercooled. If instead the liquid crystallizes, the accessible region for the determination of the fragility is restricted to temperatures between $T_m$ and the boiling (or decomposition) temperature. Then precise temperature-dependent measurements of relatively small relaxation times or low viscosities are required to derive the fragility (cf. Fig. 1) which often is experimentally challenging.

## V. SUMMARY AND CONCLUSIONS

In summary, we have shown that the thermal expansion coefficient of crystals depends in a similar way on the melting temperature as previously found for the glass-temperature dependence of the thermal expansion of glasses and supercooled liquids [8]. In particular, $\alpha_c(T_m)$ is not simply proportional to $1/T_m$ as expected when adapting the basic concepts of crystal melting that lead to the time-honored Lindemann criterion. Instead, $\alpha_c$ divided by the fragility index $m$ is well consistent with such a proportionality [Eq. (6)]. At first glance, this is surprising, because the fragility was originally introduced to classify supercooled liquids. However, as discussed in the previous section, fragility in fact affects the properties of liquids even above their melting point. As clearly revealed in Fig. 1, the viscosities of the liquids of strong and fragile glasses differ by many decades at the melting temperature. That is, crystals melting into a fragile liquid immediately attain a low-viscous state, while those transforming into a strong liquid exhibit much higher viscosity, probably corresponding to strongly different binding forces. In addition, for fragile liquids, already at $T_m$ the cooperativity of particle motion has considerably risen in relation to the single-particle motion, assumed to dominate at highest temperatures. This can be concluded from the fact that in all but the strongest liquids, close to $T_m$ the slope in Fig. 1 is already significantly larger than the slope for $1/T \rightarrow 0$. Within the nowadays quite widely accepted rationalization of non-Arrhenius behavior in terms of cooperativity, this high-temperature slope essentially reflects the energy barrier due to non-cooperative single-particle motion, because there the thermal energy far exceeds the interparticle interaction energies responsible for cooperativity. In contrast, the increasing slope and, thus, larger energy barriers at lower temperatures is caused by cooperativity, whose length scale continuously rises with decreasing temperature [18,19,27,28,36,131].

As a tentative scenario to *qualitatively* understand the approximate validity of Eq. (6), it then seems reasonable that for fragile systems the melting of a crystal not only requires the overcoming of the interparticle binding strength (directly related to the pair-potential depth $U_0$), which would lead to Eq. (1). Instead, additional thermal energy must be invested for melting because the resulting liquid is cooperative. Cooperativity leads to a reduction of configurations available to particle rearrangement, resulting in smaller entropy. In accord with the reasoning of the Adam-Gibbs theory [27], which ascribes the mentioned energy-barrier increase upon cooling to a cooperativity-induced reduction of entropy, the Gibbs free energy in fragile liquids is enhanced. Therefore, considerably more energy must be invested in order to liquify a crystal into a fragile liquid state, leading to larger $T_m$ than without cooperativity. In other words, the melting point is determined by the crossing of the temperature-dependent free energies of the crystal and liquid states [69], and cooperativity increases this energy for the liquid state via entropy reduction. This causes the melting point to rise.

To *quantitatively* understand Eq. (6), one needs to explain why cooperativity should enhance $T_m$ by just a factor $m/16$, an ad-hoc assumption made in section IV (in analogy to Ref. [8]) to rationalize this equation. We want to clearly state, that, to our knowledge, currently there is no theoretical foundation for such a proportionality. Experimentally, its validity is justified by the restoration of the $1/T_m$ dependence of $\alpha_c$ when scaling it by $m$ [Fig. 3(b)], in accord with Eq. (6). However, in view of the data scatter, currently we can only state that the available experimental data are well consistent with this relation, which implies an enhancement of $T_m$ proportional to $m$. Overall, it is clear that more theoretical and experimental work is desirable to finally clarify these issues. The purpose of the present work is to trigger such further investigations, which appear highly rewarding: A final confirmation would lead to a fundamentally different picture of crystal melting: for all materials it seems to be strongly influenced by cooperativity, a quantity usually considered to be only relevant for glass-forming liquids and the glass transition.

---

* Corresponding author.
Peter.Lunkenheimer@Physik.Uni-Augsburg.de

# Supplemental Material

for

# Crystal melting influenced by particle cooperativity of the liquid

P. Lunkenheimer[1,*], K. Samwer[2], and A. Loidl[1]

[1]*Experimental Physics V, Center for Electronic Correlations and Magnetism, University of Augsburg, 86159 Augsburg, Germany*

[2]*1. Physikalisches Institut, University of Göttingen, 37077 Göttingen, Germany*

*e-mail: peter.lunkenheimer@physik.uni-augsburg.de

TABLE SI. Glass-transition temperatures $T_g$ and melting temperatures $T_m$ of various glass formers as used for Fig. 2.

| | $T_g$ (K) | $T_m$ (K) |
|---|---|---|
| **Molecular:** | | |
| 1-Butene | 59 [1] | 88 [1] |
| Isopentane | 65 [2] | 113 [2] |
| 2,3-Dimethylbutane | 76 [1] | 136 [1] |
| 2-Methylpentane | 80 [1] | 120 [1] |
| Cyclohexene | 81 [1] | 139 [1] |
| 2-Butanethiol | 90 [1] | 133 [1] |
| 1-Propanol | 96 [3] | 146 [4] |
| Ethylcyclohexane | 98 [1] | 162 [1] |
| Ethanol | 99 [3] | 161 [4] |
| Methanol | 103 [2] | 175 [2] |
| Ethylbenzene | 111 [1] | 178 [1] |
| 4-Methylnonane | 113 [1] | 175 [1] |
| Toluene | 113 [1] | 178 [1] |
| n-Butylcyclohexane | 119 [1] | 198 [1] |
| n-Propylbenzene | 122 [1] | 174 [1] |
| Vinyl acetate | 125 [2] | 180 [2] |
| n-Butylbenzene | 125 [1] | 185 [1] |
| Isopropylbenzene | 126 [1] | 177 [1] |
| N-[β-(trimethylsilyl)ethyl]trimethylenimine | 127 [1] | 200 [1] |
| Vinyldimethylphenylsilane | 130 [1] | 191 [1] |
| $H_2SO_4$-$3H_2O$ | 158 [1] | 237 [1] |
| Propylene carbonate | 159 [3] | 224 [5] |
| Propylene glycol | 168 [3] | 214 [5] |
| Diethyl phthalate | 180 [1] | 270 [1] |
| Glycerol | 185 [3] | 291 [4] |
| *dl*-Lactic acid | 200 [4] | 291 [4] |
| Benzophenone | 212 [6] | 321 [7] |
| Salol | 218 [3] | 315 [7] |
| α-Phenyl-o-cresol | 223 [3] | 323 [5] |
| 1,1'-bis(p-methoxyphenyl)cyclohexane (BMPC) | 240 [3] | 331 [8] |
| Ortho-Terphenyl (OTP) | 245 [3] | 329 [9] |
| Xylitol | 248 [3] | 366 [10] |
| 1,1'-di(4-methoxy-5-methylphenyl)cyclohexane (BMMPC) | 261 [3] | 346 [8] |
| Sorbitol | 274 [3] | 388 [11] |
| Glucose | 305 [3] | 414 [4] |
| Sucrose | 340 [3] | 461 [12] |
| Phenolphthalein | 340 [4] | 534 [4] |
| ααβ-tris-naphthylbenzene (TNB) | 342 [3] | 470 [13] |

TABLE SI. (*continued*)

| | | |
|---|---|---|
| **Polymers:** | | |
| Polydimethylsiloxane | 150 [4] | 215 [4] |
| Polybutadiene | 188 [3] | 285 [14] |
| Polyisobutylene | 195 [3] | 317 [14] |
| Natural rubber | 200 [4] | 300 [4] |
| Polyisoprene | 200 [14] | 301 [14] |
| Poly(propylene oxide) | 201 [1] | 348 [1] |
| Poly(ethylene adipate) | 203 [4] | 323 [4] |
| Poly(tetramethylene sebacate) | 216 [4] | 337 [4] |
| Poly(ethylene oxide) | 219 [15] | 340 [15] |
| Polyethylene | 237 [16] | 415 [17] |
| Polypropylene | 259 [18] | 447 [18] |
| Poly(butyleneterephthalate) | 316 [19] | 503 [20] |
| Poly($\varepsilon$-aminocapramide) | 323 [4] | 498 [4] |
| Poly(hexamethylene adipamide) | 323 [4] | 538 [2] |
| Poly(piperazine sebacamide) | 355 [4] | 453 [4] |
| Poly(ethylene terephthalate) | 353 [4] | 543 [4] |
| poly(vinyl chloride) | 355 [3] | 546 [21] |
| Polystyrene | 365 [3] | 516 [22] |
| Poly(methyl methacrylate) | 378 [3] | 450 [14] |
| Polycarbonate | 415 [3] | 590 [23] |
| Poly(2,6-dimethylphenylene oxide) | 483 [3] | 548 [1] |
| **Ionic:** | | |
| Bmim Cl | 228 [3] | 330 [24] |
| $2Ca(NO_3)_2{:}3KNO_3$ (CKN) | 333 [3] | 480 [25] |
| $AgPO_3$ | 463 [3] | 761 [26] |
| **Metallic:** | | |
| $Mg_{65}Cu_{25}Y_{10}$ | 380 [3] | 730 [27] |
| $La_{55}Al_{25}Ni_{20}$ | 465 [3] | 704 [28] |
| $Pt_{57.3}Cu_{14.6}Ni_{5.3}P_{22.8}$ | 482 [27] | 754 [27] |
| $Pt_{45}Ni_{30}P_{25}$ | 496 [3] | 773 [29] |
| $Pd_{43}Cu_{27}Ni_{10}P_{20}$ | 568 [27] | 818 [27] |
| $Pd_{40}Ni_{40}P_{20}$ | 569 [3] | 884 [30] |
| $Pd_{40}Cu_{30}Ni_{10}P_{20}$ | 578 [27] | 798 [27] |
| $Pd_{48}Ni_{32}P_{20}$ | 580 [3] | 1016 [31] |
| $Pd_{16}Ni_{64}P_{20}$ | 591 [3] | 1010 [31] |
| $Zr_{46.75}Ti_{8.25}Cu_{7.5}Ni_{10}Be_{27.5}$ | 597 [27] | 1050 [27] |
| $Zr_{41.2}Ti_{13.8}Cu_{12.5}Ni_{10}Be_{22.5}$ | 625 [3] | 937 [27] |
| $Pd_{77.5}Cu_6Si_{16.5}$ | 636 [3] | 1020 [32] |
| $Zr_{65}Cu_{17.5}Al_{7.5}Ni_{10}$ | 653 [3] | 1180 [32] |
| $Zr_{11}Cu_{47}Ti_{34}Ni_8$ | 658 [3] | 1160 [32] |
| $Zr_{58.5}Cu_{15.6}Ni_{12.8}Al_{10.3}Nb_{2.8}$ | 666 [27] | 1109 [27] |
| **Network silicates:** | | |
| $PbSiO_3$ | 695 [2] | 1040 [2] |
| $Na_2Si_2O_5$ | 713 [3] | 1147 [33] |
| $69.0SiO_2{:}18.9Al_2O_3{:}12.3Na_2O$ wt% (Albite) | 922 [3] | 1373 [34] |
| $49.8SiO_2{:}25.6CaO{:}24.6MgO$ mol% (Diopside) | 1013 [3] | 1670 [35] |
| $55.6SIO_2{:}22.2Al_2O_3{:}22.2MgO$ mol% (Cordierite) | 1096 [3] | 1740 [35] |
| $51.1SiO_2{:}25.2Al_2O_3{:}23.8CaO$ mol% (Anorthite) | 1111 [3] | 1826 [36] |
| $SiO_2$ | 1446 [3] | 2003 [37] |

TABLE SI. (*continued*)

| | | |
|---|---|---|
| **Other network systems:** | | |
| $Na_2S_2O_3$ | 230 [4] | 321 [4] |
| S | 246 [3] | 392 [37] |
| Se | 310 [3] | 490 [37] |
| $TlAsTe_2$ | 338 [2] | 475 [2] |
| $TlAsSe_2$ | 373 [2] | 578 [2] |
| $TlAsS_2$ | 378 [2] | 578 [2] |
| $ZnCl_2$ | 380 [3] | 590 [37] |
| $As_2Te_3$ | 413 [2] | 685 [2] |
| $As_2O_3$ | 433 [2] | 585 [2] |
| AsSe | 443 [2] | 573 [2] |
| $As_2S_3$ | 444 [2] | 572 [2] |
| $As_2Se_3$ | 445 [3] | 645 [2] |
| $P_2O_5$ | 537 [2] | 853 [2] |
| $B_2O_3$ | 554 [3] | 793 [37] |
| $50P_2O_5$:$50Na_2O$ mol% ($NaPO_3$) | 563 [3] | 901 [2] |
| $BeF_2$ | 663 [3] | 821 [37] |
| $CdGeAs_2$ | 673 [2] | 973 [2] |
| $GeO_2$ | 787 [3] | 1359 [37] |

TABLE SII. Glass temperatures $T_g$, fragility indices $m$, and thermal volume expansion coefficients in the glass state $\alpha_g$ of several glass formers as used for Fig. 3(a), in addition to those materials already included in Fig. 1(e) of Ref. [3] and listed in the Supplementary Table 1 of that work.

| | $T_g$ (K) | $m$ | $10^4$ $\alpha_g$ ($K^{-1}$) |
|---|---|---|---|
| $Pt_{60}Ni_{15}P_{25}$ | 478 [3] | 50 [38] | 0.375 [3] |
| $Pt_{45}Ni_{30}P_{25}$ | 496 [3] | 48 [29] | 0.38 [3] |
| $Pd_{42.5}Ni_{7.5}Cu_{30}P_{20}$ | 525 [3] | 59 [39] | 0.534 [3] |
| $Pd_{43}Cu_{27}Ni_{10}P_{20}$ | 568 [27] | 73 [27] | 0.51 [40] |
| $Pd_{48}Ni_{32}P_{20}$ | 580 [3] | 55 [29] | 0.42 [3] |
| $Zr_{65}Cu_{17.5}Al_{7.5}Ni_{10}$ | 653 [3] | 46 [41] | 0.339 [3] |
| $55.6SIO_2$:$22.2Al_2O_3$:22.2MgO mol% (Cordierite) | 1096 [3] | 25 [42] | 0.108 [3] |

TABLE SIII. Melting temperatures $T_m$, fragility indices $m$, and thermal volume expansion coefficients in the crystal state $\alpha_c$ of various glass formers as used for Fig. 3(b).

| | $T_m$ (K) | $m$ | $10^4\ \alpha_c$ ($K^{-1}$) |
|---|---|---|---|
| **Molecular:** | | | |
| Glycerol | 291 [4] | 53 [3] | 1.5 [43] |
| Salol | 315 [7] | 73 [3] | 2.3 [44] |
| Benzophenone | 321 [7] | 125 [45] | 4.17 [46] |
| Ortho-Terphenyl (OTP) | 329 [9] | 81 [3] | 2.63 [9] |
| Xylitol | 366 [10] | 86 [3] | 0.97 [47] |
| Sorbitol | 388 [11] | 93 [3] | 1.5 [11] |
| Glucose | 414 [4] | 79 [3] | 0.84 [48] |
| ααβ-tris-naphthylbenzene (TNB) | 470 [13] | 66 [3] | 1.25 [13] |
| | | | |
| **Polymers:** | | | |
| Polypropylene (PP) | 447 [18] | 137 [16] | 3.45 [49] |
| Poly(butyleneterephthalate) | 503 [20] | 136 [19] | 5.1 [20] |
| | | | |
| **Metallic:** | | | |
| $Pt_{45}Ni_{30}P_{25}$ | 773 [29] | 48 [29] | 0.41 [50] |
| $Pd_{43}Cu_{27}Ni_{10}P_{20}$ | 818 [27] | 73 [27] | 0.44 [40] |
| $Pd_{40}Ni_{40}P_{20}$ | 884 [30] | 50 [3] | 0.39 [30] |
| $Zr_{41.2}Ti_{13.8}Cu_{12.5}Ni_{10}Be_{22.5}$ | 937 [27] | 39 [3] | 0.38 [51] |
| $Pd_{48}Ni_{32}P_{20}$ | 1016 [31] | 55 [29] | 0.46 [50] |
| $Pd_{77.5}Cu_6Si_{16.5}$ | 1020 [32] | 60 [3] | 0.36 [52] |
| $Zr_{11}Cu_{47}Ti_{34}Ni_8$ | 1160 [32] | 47 [3] | 1.0 [32] |
| $Zr_{65}Cu_{17.5}Al_{7.5}Ni_{10}$ | 1180 [32] | 46 [41] | 0.51 [32] |
| | | | |
| **Network silicates:** | | | |
| $69.0SiO_2$:$18.9Al_2O_3$:$12.3Na_2O$ wt% (Albite) | 1373 [34] | 22 [53][a] | 0.48 [54] |
| $49.8SiO_2$:25.6CaO:24.6MgO mol% (Diopside) | 1670 [35] | 53 [3] | 0.33 [55] |
| $55.6SIO_2$:$22.2Al_2O_3$:22.2MgO mol% (Cordierite) | 1740 [35] | 25 [42] | 0.054 [56] |
| $51.1SiO_2$:$25.2Al_2O_3$:23.8CaO mol% (Anorthite) | 1826 [36] | 54 [3] | 0.11 [57] |
| | | | |
| **Other network systems:** | | | |
| S | 392 [37] | 86 [3] | 2.1 [37] |
| Se | 490 [37] | 87 [3] | 1.29 [37] |
| $GeO_2$ | 1359 [37] | 20 [3] | 0.19 [37] |

[a]In Fig. 1 of Ref. 3, we erroneously used $m$ = 26 instead of 22 for this material, wich does not affect the conclusions.

---